\documentstyle[aps,epsfig,prl]{revtex}
\begin{document}
\title{Profile alterations of a symmetrical light pulse coming
through a quantum well}
\author{ L. I. Korovin, I. G. Lang}
\address{A. F. Ioffe Physical-Technical Institute, Russian
Academy of Sciences, 194021 St. Petersburg, Russia}
\author{D. A. Contreras-Solorio, S. T. Pavlov\cite{byline1}}
\address {Escuela de Fisica de la UAZ, Apartado Postal C-580,
98060 Zacatecas, Zac., Mexico}
\twocolumn[\hsize\textwidth\columnwidth\hsize\csname
@twocolumnfalse\endcsname
\date{\today}
\maketitle \widetext
\begin{abstract}
\begin{center}
\parbox{6in} {The theory of a response of a two-energy-level
system, irradiated by symmetrical light pulses, has been
developed.(Suchlike electronic system approximates under the
definite conditions a single ideal quantum well (QW) in a strong
magnetic field ${\bf H}$, directed perpendicularly to the QW's
plane, or in magnetic field absence.) The  ground state system
energy level is the first energy level, the excitation discrete
energy level with the energy $\hbar\omega_0$ (for instance, an
excitonic energy level at $H=0$ or any energy level in a strong
magnetic field) is the second energy level. It is supposed that
one can neglect a light-lattice interaction and influence of all
other energy levels. The general formulae for the time-dependence
of non-dimensional reflection ${\cal R}(t)$, absorption ${\cal
A}(t)$ and transmission ${\cal T}(t)$ of  a symmetrical light
pulse have been obtained. It has been shown that singularities of
three types exist on the dependencies ${\cal R}(t)$, ${\cal
A}(t)$, ${\cal T}(t)$. In the first type of singularity $t_0$
${\cal R}(t)={\cal R}(t)=0$ and the total reflection is realized.
In the case $\gamma_r>>\gamma$ ($\gamma_r$ is the radiative
lifetime broadening, $\gamma$ is the non-radiative lifetime
broadening) and under the resonant condition $\omega_l=\omega_0$
the strong alterations of the value and profile of the transmitted
pulse can happen. In the case of the long pulse
($\gamma_r>>\gamma_l$, $\gamma_l$ is the lifetime broadening of
the exciting pulse) it is almost totally reflected. In the case of
the intermediate pulse
 $(\gamma_r\simeq\gamma_l)$ reflection, absorption and
transmission are comparable in values, the transmitted pulse
profile distinguishes  strongly from the exciting pulse profile:
the transmitted pulse has two maxima due to the total reflection
point $t_0$, when transmission is absent. The oscillating time
dependence of ${\cal R}(t)$, ${\cal A}(t)$, ${\cal T}(t)$ on the
detuning frequency $\Delta\omega=\omega_l-\omega_0$ takes place.
The oscillations are better observable when
$\Delta\omega\simeq\gamma_l$. The positions of the total
absorption, reflection and transparency singularities are examined
when the frequency $\omega_l$ is detuned.}
\end{center}
\end{abstract}

] \narrowtext

\section{Introduction.}

The strong alteration of the profile of the sharply asymmetrical
light pulse, coming through a QW, and the large value of the
reflected pulse have been predicted earlier\cite{1}. It has been
supposed that the carrier frequency of the exciting pulse
$\omega_l$ is close to the electronic excitation frequency
$\omega_0$ measured from the ground state frequency. These
phenomena are possible under condition
\begin{equation}
\label{1} \gamma_r>>\gamma ,
\end{equation}
It is well known, that under the opposite condition
\begin{equation}
\label{2} \gamma_r<<\gamma
\end{equation}
the profile of the transmitted pulse changes weakly and the
reflected and absorbed pulses are weak comparatively to the
exciting pulse.

The radiative lifetime broadening of the energy levels appears in
the quasi-2D systems due to a translation  symmetry violation into
the direction perpendicular to the QW plane\cite{2,3}. In the
perfect QWs the radiative lifetime broadening
 $\gamma_r$ may be comparable to and even exceed the contributions
 of other relaxation mechanisms.
This new physical situation requires the adequate theoretical
description where the upper orders of the electron-EMF (electro
magnetic field) are taken into
account\cite{1,2,3,4,5,6,7,8,9,10,11,12,13,14}. $\gamma_r$  has
been calculated for an excitonic energy level in a QW at $H=0$ in
\cite{2}, in a strong magnetic field in\cite{11,12}, for a
magnetopolaron in a QW in\cite{12}, respectively.

 In this
article the case of the two energy level system, consisting of the
ground state and an excited energy level, has been considered.
Influence of other energy levels is neglected. An exciton in a QW
in a strong (zero) magnetic field is considered as an electronic
excitation. The case of the symmetrical pulse is considered unlike
 \cite{1}, because
 a pulse with a sharp front is  very difficult to realize.

\section{Electric fields on the right and left side of a QW,
 irradiated by light pulses.}

Let us suppose that from the left (where $z<0$) an exciting light
pulse with the corresponding electric field
 \begin{eqnarray}
\label{3} {\bf E}_0(z,t)&=&E_0{\bf e}_le^{-i\omega_l p}\nonumber\\
&\times&\{\Theta (p)e^{-\gamma_{{l1}}p/2}\
+[1-\Theta(p)]e^{\gamma_{{l2}}p/2}\} + c. c.,
\end{eqnarray}
incidents on the single QW ($E_0$ is the real amplitude, ${\bf
e}_l$ is the polarization vector,
 $p=t-zn/c$, n is the refraction index out the QW, $\Theta(p)$
 is the Haeviside step-function, $\gamma_{l1} (\gamma_{l2})$
  determine a damping (increasing) of
 the symmetrical light pulse). The Umov-Poynting vector
 \begin{equation}
\label{4} {\bf S}(p)={\bf S}_{0}P(p),
\end{equation}
  \begin{eqnarray}
\label{5} {\bf S}_{0}={\bf e}{_z}cE_0^2/(2\pi n),\nonumber\\
P(p)=\Theta (p)e^{-\gamma_
{{l1}}p/2}+[1-\Theta(p)]e^{\gamma_{{l2}}p/2},
\end{eqnarray}
corresponds to the pulse of Eq. (3), ${\bf e}_z$ is the unit
vector along the $z$ axis.

After the Fourier transform  Eq. (3) takes the form
 \begin{equation}
\label{6} {\bf E}_0(z,t)=E_0{\bf e}_l\int_{-\infty}^{\infty}
d\omega e^{-i\omega p} {\cal D}_0(\omega)+ c. c. ,
\end{equation}
where
 \begin{equation}
\label{7} {\cal
D}_0(\omega)=\frac{i}{2\pi}\left[{1\over\omega-\omega_l+i\gamma_{l1}/2}-
{1\over\omega-\omega_l-i\gamma_{l2}/2)}\right].
\end{equation}

In \onlinecite{1,10,11} the strongly asymmetrical pulse has been
used with a sharp front, for which $\gamma_{l2}\to\infty$ and the
second term in Eq. (5) vanishes, as well as the second term in the
square brackets in Eq. (7). At
 \begin{equation}
\label{8} \gamma_{l1}=\gamma_{l2}=\gamma_{l}
\end{equation}
the pulse Eq. (3) is symmetrical. At $\gamma_l \to 0$ the
symmetrical pulse transfers into the monochromatic wave with the
frequency $\omega_l$,  and the function ${\cal D}_o(\omega)$
transfers into $\delta (\omega-\omega_l)$. The pulse Eq. (3) is
very useful for calculations. Its defect is the breakdown of the
derivative in the point $t-zn/c$ (see Eq. (5)), however all the
qualitative conclusions of our theory below do not change when one
transfers to the smooth pulses. Some results  for the symmetrical
pulse proportional to $1/\cosh{(\gamma_l p)}$ have been
demonstrated in \onlinecite{1}.

Let us consider QWs with the width $d$ which is  smaller than the
light wave length
 $c/(n\omega_l)$. Then the electric fields ${\bf E}_{left(right)}(z,t)$
on the left (right) side of the QW are determined by the
expressions \cite{1}
 \begin{equation}
\label{9} {\bf E}_{left(right)}(z,t)={\bf E}_0(z,t)+\Delta {\bf
E}_{left(right)}(z,t),
\end{equation}
\begin{eqnarray}
\label{10} \Delta {\bf E}_{left(right)}(z,t)=E_0{\bf
e}_l\nonumber\\ \times\int_{-\infty}^{\infty}
 d\omega e^{-i\omega (t\pm zn/c)}
{\cal D}(\omega)+ c. c. ,
\end{eqnarray}
where the upper (lower) sign refers to $left(right)$. According to
Eq. (10), the polarization of the induced electric field coincides
with the exciting field polarization. The result of Eq. (10)
supposes that the wave has a circular polarization
\begin{equation}
\label{11} {\bf e}_l=({\bf e}_x+i{\bf e}_y)/\sqrt{2},
\end{equation}
where ${\bf e}_x$, ${\bf e}_y$ are the unit vectors along the axis
$x$ and $y$, respectively. It is implied also that each of the
circular polarizations corresponds to the excitation  from the
ground state of one of two types of the electron-hole pairs (EHPs)
with equal energies (see \onlinecite{12,16}).

The frequency dependence  ${\cal D}(\omega)$ is as follows
\begin{equation}
\label{12} {\cal D}(\omega)=-\frac{4\pi \chi(\omega){\cal
D}_0(\omega)}{1+ 4\pi \chi (\omega)},
\end{equation}
\begin{eqnarray}
\label{13} \chi(\omega)=(i/4\pi)\sum_\varrho
(\gamma_{r\varrho}/2)[(\omega-
\omega_{\varrho}+i\gamma_{\varrho}/2)^{-1}\nonumber\\ +(\omega+
\omega_{\varrho}+i\gamma_{\varrho}/2)^{-1}],
\end{eqnarray}
where $\varrho$ is the number of the excited state,
$\hbar\omega_\varrho$ is the energy of the excited state, measured
from the ground state energy, $\gamma_{r\varrho}$ (
$\gamma_\varrho$) is the radiative (non-radiative) lifetime
broadening of the excited state
 $\varrho$. The second term in the square brackets of
Eq. (13) is non-resonant one and neglected below. It is implied
also, that the light reflection and light absorption by the QW is
due to the electron transitions from the valence band into the
conductivity band. The light-lattice interaction as well as the
interaction with the deep energy levels are neglected.

As it was noted above we will take into account in the sum of Eq.
(13) the only excited energy level, i. e. we consider the
two-level system, where the first energy level corresponds to the
ground state and the second corresponds to the excited energy
level. The index $\varrho$ takes only one value, therefore we use
the designations:
\begin{equation}
\label{14}
\omega_\varrho=\omega_0,\quad\gamma_{r\varrho}=\gamma_r,
\quad\gamma_\varrho=\gamma,\quad\Gamma=\gamma_r+\gamma.
\end{equation}
With the help of Eqs. (7) - (13) we obtain the following result
for the induced field on the left side of the QW when irradiated
by the symmetrical pulse
 \begin{eqnarray}
\label{15} \Delta {\bf E}_{left}(z,t)=-iE_0{\bf
e}_l(\gamma_r/2)\nonumber\\\times\left\{{\cal E}
\Theta(s)+\frac{e^{-i\omega_0s-\Gamma
s/2}[1-\Theta(s)]}{\Delta\omega+i(\Gamma +\gamma{_l})/2}\right\}+
c. c.,
\end{eqnarray}
$${\cal
E}=\left\{\frac{e^{-i\omega_{l}s-\gamma_{l}s/2}}{\Delta\omega+i(\Gamma
-\gamma{_l})/2}\right.$$ $$-e^{-i\omega_0s-\Gamma
s/2}\left.\left[\frac{1}{\Delta\omega+i(\Gamma
+\gamma{_l})/2}\right]\right\},$$ where $s=t+zn/c$,
$\Delta\omega=\omega_l-\omega_0$. The expression
 $\Delta {\bf E}_{right}(z,t)$ distinguishes from Eq. (15)
 only by the substitution  
 $p=t-zn/c$ instead of $s$. Because Eq. (15) is right for  $|z|>>d$,
we neglect the QW width below and believe, that the QW is in the
plane  $z=0$.

\section{Transmitted, reflected and absorbed
 energy fluxes.}

For the sake of brevity let us call the Umov-Poynting vector as
the energy flux. The transmitting flux (i. e. the flux on the
right of the QW) is equal
\begin{equation}
\label{16} {\bf S}_{right}(z,t)=({\bf e}_z/4\pi)(c/n)({\bf
E}_{right}(p))^2,
\end{equation}
the flux on the left side of the QW is
 \begin{equation}
\label{17} {\bf S}_{left}(z,t)={\bf S}(p)+{\bf S}_{ref}(s),
\end{equation}
where ${\bf S}(p)$ is the exciting pulse flux, determined in Eq.
(4),
 ${\bf S}_{ref}(s)$ is the reflected flux:
\begin{equation}
\label{18} {\bf S}_{ref}(s)=-({\bf e}_z/4\pi)(c/n)(\Delta{\bf
E}_{left}(s))^2.
\end{equation}
The absorbed energy flux is determined as
 \begin{equation}
\label{19} {\bf S}_{abs}(t)={\bf S}_{left}(z=0,t)-{\bf
S}_{right}(z=0,t)
\end{equation}
and equals
 \begin{equation}
\label{20} {\bf S}_{abs}(t)=-({\bf e}_z/2\pi)(c/n){\bf
E}_{right}(z=0,t) \Delta{\bf E}(t),
\end{equation}
where
 \begin{equation}
\label{21} \Delta{\bf E}(t)=\Delta{\bf E}_{left}(z=0,t)=\Delta{\bf
E}_{right}(z=0,t).
\end{equation}
Let us introduce the non-dimensional functions of transmission
${\cal T}(x)$, reflection ${\cal R}(x)$ and absorption ${\cal
A}(x)$, determining them as
 \begin{eqnarray}
\label{22} {\bf S}_{right}(z,t)={\bf S}_0{\cal T}(p),\quad {\bf
S}_{ref}(z,t)=-{\bf S}_0{\cal R}(s),\nonumber\\ {\bf S}_{abs}(t)=
{\bf S}_0{\cal A}(t).
\end{eqnarray}
It follows from Eq. (19), that
 \begin{equation}
\label{23} {\cal T}(x)+{\cal R}(x)+{\cal A}(x)=P(x).
\end{equation}
The values $P(x)$, ${\cal T}(x)$ and ${\cal R}(x)$ are always
positive, absorption may be positive, as well as negative. The
negative absorption means that at some moment  $t$ the QW
electronic system gives back the accumulated energy. Below we use
the designation $t$ instead of $x$ remembering the definitions of
Eq. (22). The variable $t$ corresponds to the real time at $z=0$.

\section{Time points of zero absorption, total reflection
 and total transparency.}

 The content of this section is based on the analysis the formulae
of the section III without specifying the expression for the
induced field $\Delta{\bf E}(z,t)$. Therefore the results obtained
above
 are justified for any number of the excited energy levels
in a QW (see for instance \onlinecite{10,11,12,13,14}), but not
only in the case of the only energy level, where the expression
Eq. (15) is applicable.

In the figures, demonstrating the curves ${\cal A}(t),$ ${\cal
R}(t)$ and ${\cal T}(t)$, one sees the points of zero absorption
and total transparency  ${\cal T}(t)=P(t)$. Let us call them
specific time points. It follows from the Eqs. (16), (18) and (20)
that there are three types of such points.

In the case of the first and second type, they correspond to
vanishing of the electric fields or their combinations. In the
case of the first type specific points  the field ${\bf
E}_{right}(z=0,t)$ or $\Delta{\bf E}_{left}(z=0,t)$ vanishes. In
the case ${\bf E}_{right}(z=0,t_0)=0$ we have
\begin{equation}
\label{24} {\cal A}(t_0)=0,\quad{\cal T}(t_0)=0,\quad{\cal
R}(t_0)=P(t_0);
\end{equation}
we call the point $t_0$ the first type total reflection point. In
the case $\Delta{\bf E}_{left}(z=0,t_x)=0$
\begin{equation}
\label{25} {\cal A}(t_x)=0,\quad{\cal R}(t_x)=0,\quad{\cal
T}(t_x)=P(t_x);
\end{equation}
we call the point  $t_x$ as the first type total transparency
point. Thus, ${\cal A}(t)=0$ in the first type specific points of
total reflection and total transparency.

The equation $P(t)-{\cal R}(t)=0$ may be written as
\begin{eqnarray}
\label{26} ({\bf E}_0)^2-(\Delta{\bf E})^2=0,\nonumber\\({\bf
E}_0+\Delta{\bf E})({\bf E}_0- \Delta{\bf E})=0,\nonumber\\{\bf
E}_{right}({\bf E}_0-\Delta{\bf E})=0,
\end{eqnarray}
and the equation  $P(t)-{\cal T}(t)=0$ may be written as
\begin{eqnarray}
\label{27} ({\bf E}_0)^2-({\bf E}_{right})^2=0,\nonumber\\({\bf
E}_0-{\bf E}_{right})({\bf E}_0+ {\bf
E}_{right})=0,\nonumber\\{\Delta\bf E}({\bf E}_0+{\bf
E}_{right})=0.
\end{eqnarray}

 Therefore
the condition ${\bf E}_0-{\Delta\bf E}=0$ corresponds to the total
reflection point and the condition ${\bf E}_0+{\bf E}_{right}=0$
corresponds to the total transparency point. Generally speaking
the absorption ${\cal A}(t)$ does not equal 0 in these points. We
call these points as the second type total reflection and total
transparency points. Finally, it follows from the Eqs. (20), (26)
and (27) that the specific points
 ${\cal A}=0, {\cal R}=P$ and ${\cal T}=P$
may appear in the case of the perpendicular vectors ${\bf
E}_{right}$ and $\Delta{\bf E}$,
 ${\bf E}_{right}$ and ${\bf E}_0-\Delta{\bf E}$, $\Delta{\bf
 E}$ and
 ${\bf E}_0+{\bf E}_{right}$, respectively. We call these specific
 points the third type specific points.

 In experiments these specific points are observable as following.
Transmission as a function of time $t$ is measured in the plane
$z_0$, reflection is in the plane  $-z_0$ , i. e. on the left of
the QW. Let us suppose, that the point  $t_0$ is the first type
total reflection point. Taking into account Eq. (22), we find,
that at the moment
 $t=t_0+z_0n/c$ one has zero transmission and total reflection.  

We cannot measure directly the absorption  ${\cal A},$ but,
determining it as  ${\cal A}=P-{\cal R}-{\cal T}$ at the time
moment $t=t_0+z_0n/c$, we obtain ${\cal A}=0$ in the case of the
first type specific point. It means that the described experiment
absorption is measured with some delay
 $\Delta t=z_0n/c$.
\section{Energy fluxes in the sharp resonance at the arbitrary
relationship between the lifetime broadenings.}

With the help of the results of section III and Eqs.
 (3), (8), (9) and (15) for the electric fields one can determine
 the values of the transmitted, reflected and absorbed energy fluxes
 at any $\omega_l$ and
 $\gamma_l$, characterizing the exciting pulse, and the parameters
$\omega_0$, $\gamma_r$ and $\gamma$, characterizing the energy
level in the QW.

At $\gamma_l=0$, which corresponds to the monochromatic
irradiation, we obtain the expressions
\begin{eqnarray}
\label{28} {\cal
T}=\frac{(\Delta\omega)^2+\gamma^2/4}{(\Delta\omega)^2+\Gamma^2/4},\nonumber\\
{\cal R}=\frac{\gamma^2/4}
{(\Delta\omega)^2+\Gamma^2/4},\nonumber\\ {\cal
A}=\frac{\gamma\gamma_r/2}{(\Delta\omega)^2+\Gamma/4},
\end{eqnarray}
 obtained earlier in \onlinecite{2,3,4,5,8}.

In the sharp resonance
\begin{equation}
\label{29} \omega_l=\omega_0
\end{equation}
we obtain the following results for the values, characterizing the
energy fluxes:
\begin{eqnarray}
\label{30} {\cal T}(p)&=&\Theta(p)[{\cal F}_T(p)]^2\nonumber\\&+&
[1-\Theta(p)]e^{\gamma_lp}(\gamma+\gamma_l)^2/(\Gamma+\gamma_l)^2,
\end{eqnarray}
$${\cal
F}_T(p)=e^{-\gamma_lp/2}(\gamma-\gamma_l)/(\Gamma-\gamma_l)
+e^{-\Gamma p/2}\gamma_rG,$$
\begin{eqnarray}
\label{31} {\cal R}(s)=\gamma_r^2\{\Theta(s)[{\cal
F}_R(s)]^2\nonumber\\+[1-\Theta(s)]
e^{\gamma_ls}(\Gamma+\gamma_l)^{-2}\},
\end{eqnarray}
$${\cal F}_R(s)={e^{\gamma_ls/2}-e^{-\Gamma
s/2}\over\Gamma-\gamma_l}+ {e^{-\Gamma
s/2}\over\Gamma+\gamma_l},$$
\begin{eqnarray}
\label{32} {\cal A}(t)&=&2\gamma_r\{\Theta(t){\cal
F}_A(t)\nonumber\\&+&[1-\Theta(t)]
e^{\gamma_lt}(\gamma+\gamma_l)/(\Gamma+\gamma_l)^2\},
\end{eqnarray}
$${\cal F}_A(t)=e^{-\gamma_lt}\frac{\gamma-\gamma_l}
{(\Gamma-\gamma_l)^2}$$ $$ -\gamma_rG^2e^{-{\Gamma
t}}+Ge^{-(\Gamma+\gamma_l)t/2}\frac{\gamma_r-\gamma+\gamma_l}
{\Gamma-\gamma_l}.$$ In Eqs (30) and(32) the designation
$$G=(\Gamma-\gamma_l)^{-1}- (\Gamma+\gamma_l)^{-1}$$ is
introduced. It follows from Eqs. (9),(15) at $\omega_l=\omega_0$,
that the position
  of the first type total reflection point $t_0$
is determined by the condition
 $t> 0$ and by the equation
\begin{equation}
\label{33}
e^{-\gamma_lt/2}\frac{\Gamma-\gamma_l-\gamma_r}{\Gamma-\gamma_l}
+e^{-\Gamma t/2}\frac{2\gamma_r\gamma_l}{\Gamma^2-\gamma_l^2}=0.
\end{equation}
The solution of Eq. (33) is as follows
\begin{equation}
\label{34}
t_0=\frac{2}{\Gamma-\gamma_l}ln\frac{2\gamma_r\gamma_l}{(\gamma_l-\gamma)
(\Gamma+\gamma_l)},\quad\gamma_l>\gamma,
\end{equation}
thus, the total reflection point always exists for the short
pulse. It means, that at large times, absorption by the QW is
substituted by generation, because ${\cal A}(t_0)=0$ and it is
negative at $t>t_0$.

The position of the second type total transparency point $t_t$,
where ${\bf E}_{right}+{\bf E}_0=0$, is determined by the
conditions
\begin{eqnarray}
\label{35} t>0,\quad
e^{-\gamma_lt/2}\left(2-\frac{\gamma_r}{\Gamma-\gamma_l}\right)
\nonumber\\+ e^{-\Gamma
t/2}\frac{2\gamma_r\gamma_l}{\Gamma^2-\gamma_l^2}=0,
\end{eqnarray}
from which we obtain
\begin{eqnarray}
\label{36}
t_t=\frac{2}{\Gamma-\gamma_l}ln\frac{2\gamma_r\gamma_l}{(\Gamma+\gamma_l)
(2\gamma_l-\Gamma-\gamma)},\nonumber\\
 \gamma_l>\gamma+\gamma_r/2.
\end{eqnarray}
According to Eq. (36), the point $t_t$  exists in the case of a
short pulse. Note that the condition $\gamma_l>\gamma+\gamma_r/2$
is harder than the condition $\gamma_l>\gamma$ in Eq. (34).

The conditions  $\Delta{\bf E} =0, {\bf E}_0-\Delta{\bf E}=0$ are
not performed for the examined case of only excited energy level
in the resonance $\omega_l=\omega_0$, therefore there are neither
the first type total transparency points nor the second type total
reflection points (see Figs. 1-3).

The third type specific time points do not exist in the resonance
either. Indeed, the vectors ${\bf E}_0$, ${\bf E}_{right}$ and
 $\Delta{\bf E}$ are parallel always.
This  can be checked with the help of Eqs. (3), (9) and(15) at
$\Delta\omega=0$. For the vectors
 ${\bf E}_0(z=0,t)$,  ${\bf E}_{right}(z=0,t)$
and $\Delta{\bf E}(z=0,t)$ and for their combinations ${\bf
E}_0-\Delta{\bf E}$ and ${\bf E}_0+{\bf E}_{right}$ we obtain  
\begin{equation}
\label{37} {\bf E}_i(z=0,t)=E_{0}F_i(t){\bf e}_l^{(\pm)}(t),
\end{equation}
where
\begin{eqnarray}
\label{38} {\bf e}_l^{(\pm)}(t)&=&{\bf e}_le^{-i\omega_lt}+{\bf
e}_l^{\ast}e^{i\omega_lt}\nonumber\\&=& \sqrt 2({\bf
e}_xcos\omega_lt\pm{\bf e}_ysin\omega_lt),
\end{eqnarray}
the sign $+(-)$ corresponds to the right (left) circular
polarization, $F_i(t)$ are the non-dimensional real functions. It
follows from Eq. (37) that three vectors are parallel. The
equalities  $F_i(t)=0$ at $t>0$ correspond to the first and second
type points.

The following picture exists for the short pulses under condition
$\gamma_l>\Gamma$ at $p>>\gamma_l^{-1}$. The exciting field, which
contains the factor $\exp(-\gamma_lp/2)$, becomes negligibly
small, therefore
\begin{equation}
\label{39} {\bf E}_{right}(z,t)\simeq\Delta{\bf E}_{right}(z,t).
\end{equation}
At $t>>\gamma_l^{-1}$ we obtain
\begin{equation}
\label{40} {\cal R}(t)\simeq{\cal T}(t),\quad{\cal
A}(t)\simeq-2{\cal R}(t).
\end{equation}
It means, that only induced fields
 $\Delta{\bf E}_{left(right)}$, symmetrical in relation to the QW
 plane, are preserved.

The electronic system gives back the accumulated energy, radiating
it symmetrically by the left and right fluxes. In Fig. 1 one sees
the fulfillment of the relations Eq. (40) at $\gamma_lt>>1$.
\vspace{0.5cm}
\section{Energy fluxes in the sharp resonance under condition $\gamma_r
>>\gamma.$}
The perturbation theory is applicable in the case
$\gamma_r<<\gamma$. It is enough to take into account only the
lowest order on the electron-EMF interaction, which is equivalent
to neglecting the term   $4\pi\chi(\omega)$ in the denominator of
Eq. (12).\cite{byline2}

 Under condition $\gamma_r<<\gamma$ the induced fields $\Delta{\bf
E}_{left(right)}$ on the left  (right) of the QW are small in
comparison to the exciting field
 Eq. (3) at $p=0$, i. e. the non-dimensional reflection $\cal
 R$ and
absorption  $\cal A$ are small in comparison to unity. The
transmitting pulse is distinguished weakly in its profile from the
exciting pulse. But even in this situation, very interesting
results have been obtained: some delaying of the short pulse is
seen in the transmitting light at the times of order
  $\gamma^{-1}$, and the sinusoidal beats on the frequency
 $\Delta E/\hbar$ are seen in the case of the closely displaced
 energy levels, where
 $\Delta E$ is the energy distance between the levels (see, for instance,
 \onlinecite{15}).

In the opposite case  $\gamma_r>>\gamma$  the induced fields are
comparable in values to the exciting fields, and the profile of
the transmitting  pulse may change drastically. It has been shown
in \onlinecite{1} for the asymmetrical pulse, and the results of
the numerical calculations for the symmetrical pulse, proportional
to $1/\cosh(\gamma_lp),$ have been represented. The analytical
expressions for the non-dimensional values $\cal T,\cal R$ and
$\cal A$ for the symmetrical pulse Eq. (3) have been represented
above in Eqs. (30) - (32).

Let us consider the case $\gamma =0$, where the condition
 $\gamma_r>>\gamma$ is  always satisfied.
 The curves, calculated with the help of Eqs.
(30) - (32) and at $\gamma =0$, are represented in Figs. 1-3 for
the short pulse $(\gamma_l>>\gamma_r)$, for the long pulse
$(\gamma_l<<\gamma_r)$ and for the intermediate pulse
$(\gamma_l=\gamma_r).$ The first type total reflection point $t_0$
is seen in figures. It always exists at  $\gamma =0$ and is
determined by the expression
\begin{equation}
\label{41}
t_0=\frac{2}{\gamma_r-\gamma_l}ln\frac{2\gamma_r}{\gamma_r+\gamma_l}.
\end{equation}
Therefore the curve ${\cal T}(t)$ has two maxima, which is seen
especially well in Fig. 3. 
 The second type total transparency point $t_t$
 is seen in Figs. 1, 3. At $\gamma =0$ it exists at $\gamma_l>\gamma_r/2$
and equals
\begin{equation}
\label{42}
t_t=\frac{2}{\gamma_r-\gamma_l}ln\frac{2\gamma_r\gamma_l}{(\gamma_r+\gamma_l)
(2\gamma_l-\gamma_r)}.
\end{equation}
In Fig. 2 this point disappears, because the condition
$\gamma_l>\gamma_r/2$ is not fulfilled for the long pulses.

It is seen from the Eqs. (30) - (32) and Fig. 1 that in the case
of the short pulse the transmitted pulse distinguishes from the
exciting pulse not very strongly. The reflected pulse is very
weak, because Eq. (31) contains the small factor
$\sim(\gamma_r/\gamma_l)^2$. Absorption is small also in
comparison to the exciting pulse, but it is larger than
reflection, because it contains the factor
 $\gamma_r/\gamma_l$.

We have a different picture in the case of the long pulse. The
pulse is almost totally reflected (see Fig. 2) and almost
coincides with the exciting pulse. The transmitted pulse is very
small, it contains the factor
 $\sim(\gamma_r/\gamma_l)^2$. Absorption is larger than
 transmission, because it contains the factor
 $\gamma_r/\gamma_l$.

Fig. 3 relates to the case $\gamma_l=\gamma_r$. Substituting
   $\gamma_l=\gamma_r$ and $\gamma=0$  in Eqs. (30) - (32) we
   obtain
$${\cal T}(t)=[\Theta (t)e^{-\gamma_lt}(1-\gamma_lt)^2+(1-\Theta
(t)) e^{\gamma_lt}]/4, $$
\begin{equation}
\label{43} {\cal R}(t)=[\Theta
(t)e^{-\gamma_lt}(1+\gamma_lt)^2+(1-\Theta (t)) e^{\gamma_lt}]/4,
\end{equation}
$${\cal
A}(t)=[\Theta(t)e^{-\gamma_lt}(1-(\gamma_lt)^2)+(1-\Theta(t))
e^{\gamma_lt}]/2.$$ Obviously, that the first type total
reflection point $t_0$ and the second type total transparency
point $t_t$ are equal
\begin{equation}
\label{44} t_0=\gamma_l^{-1},\qquad t_t=3\gamma_l^{-1}.
\end{equation}

The exciting, transmitting and reflected pulses are of the same
order values, but the transmitting pulse changes strongly in
profile from the exciting pulse, which is seen in Fig. 3. In the
point $t_0=\gamma_l^{-1}$ the transmitting pulse has a minimum,
afterwards it has a second maximum.

\section{Integral energy fluxes in the resonance
$\omega_l=\omega_0.$}

At the pulse irradiation the total energy, absorbed on a unit
area, is as follows
\begin{equation}
\label{45} {\cal E_A}=2|S_0|{\cal K}_{\cal A}/\gamma_l,
\end{equation}
where the non-dimensional value is
\begin{equation}
\label{46} {\cal K}_{\cal
A}=(\gamma_l/2)\int_{-\infty}^{\infty}{\cal A}(t)dt.
\end{equation}
Reciprocally, the total exciting, transmitted and reflected
energies on a unit  area are as follows
\begin{eqnarray}
\label{47} {\cal E}_P=2|S_0|{\cal K}_P/\gamma_l,\quad{\cal
E}_{\cal T}= 2|S_0|{\cal K}_{\cal T}/\gamma_l,\nonumber\\{\cal
E}_{\cal R}= 2|S_0|{\cal K}_{\cal R}/\gamma_l,
\end{eqnarray}

where
\begin{eqnarray}
\label{48} {\cal
K}_P&=&\frac{\gamma_l}{2}\int_{-\infty}^{\infty}P(t)dt, {\cal
K}_{\cal
T}\nonumber\\&=&\frac{\gamma_l}{2}\int_{-\infty}^{\infty}{\cal
T}(t)dt, {\cal K}_{\cal
R}\nonumber\\&=&\frac{\gamma_l}{2}\int_{-\infty}^{\infty}{\cal
R}(t)dt,
\end{eqnarray}
with
\begin{equation}
\label{49} {\cal K}_{\cal T}+{\cal K}_{\cal R}+{\cal K}_{\cal
A}={\cal K}_P.
\end{equation}
With the help of Eqs. (30) - (32) we obtain
\begin{equation}
\label{50} {\cal K}_{\cal
T}=\frac{\gamma^2(\Gamma+2\gamma_l)+\gamma_l^2\Gamma}
{\Gamma(\Gamma+\gamma_l)^2},
\end{equation}
\begin{equation}
\label{51} {\cal K}_{\cal
R}=\frac{\gamma_r^2(\Gamma+2\gamma_l)}{\Gamma(\Gamma+
\gamma_l)^2},
\end{equation}
\begin{equation}
\label{52} {\cal K}_{\cal
A}=\frac{2\gamma_r\gamma(\Gamma+2\gamma_l)}{\Gamma(\Gamma+
\gamma_l)^2}.
\end{equation}
The analogous  expressions for the asymmetrical pulse have been
obtained in \onlinecite{1}. It follows from Eq. (52), that at
$\gamma=0$ the total absorbed energy equals 0. This is clear
physically, because only at $\gamma\neq 0$ the electronic
excitations in a QW transfer the energy to other excitations, for
instance, to phonons. If $\gamma=0$, the whole energy of the
electronic excitations transfers into the light energy, and
absorption vanishes. Suppose, that $\gamma=0, \gamma_r=\gamma_l$,
then we obtain from Eq. (50) and Eq. (51) that three quarters of
the exciting pulse energy are reflected, and one quarters goes
through the QW.

\section{Light pulse reflection and absorption  when the carrier
 frequency
is detuned.}

The expressions for the non-dimensional absorption ${\cal A}(t)$,
reflection ${\cal R}(t)$ and transmission  ${\cal T}(t)$ are as
follows
\begin{eqnarray}
\label{53} {\cal A}(t)&=&\frac{\Theta
(t)}{(\Delta\omega)^2+(\gamma_l-\Gamma)^2/4}
\nonumber\\&\times&\left\{\frac{e^{-\gamma_lt}
\gamma_r(\gamma-\gamma_l)}{2}- \frac{e^{-\Gamma
t}\gamma_r^2\gamma_l^2}
{2[(\Delta\omega)^2+(\gamma_l+\Gamma)^2/4]}\right.\nonumber\\
&-&\gamma_r\gamma_l\sqrt\frac{(\Delta\omega)^2+(\gamma_l+\gamma_r-
\gamma)^2/4}{(\Delta\omega)^2+(\gamma_l+\Gamma)^2/4}\nonumber\\
&\times&\left. e^{-(\gamma_l+\Gamma)t/2}\cos(\Delta\omega
t+\chi)\right\}\nonumber\\&+&[1-\Theta
(t)]\frac{e^{\gamma_lt}\gamma_r(\gamma_l+\gamma)}
{(\Delta\omega)^2+(\gamma_l+\Gamma)^2/4},
\end{eqnarray}
\begin{eqnarray}
\label{54} {\cal R}(t)&=&\frac{\gamma_r^2}{4}\left\{\frac{\Theta
(t)} {(\Delta\omega)^2+(\gamma_l-\Gamma)^2/4}\right.\nonumber\\
&\times&\left [a_{\cal R}^{2}(t)+ b_{\cal R}^{2}(t)+ +2a_{\cal
R}(t)b_{\cal R}(t)\cos(\Delta\omega t-\zeta)\right]\nonumber\\
&+&\left. \frac{[1-\Theta
(t)]e^{\gamma_lt}}{(\Delta\omega)^2+(\gamma_l+\Gamma)^2/4}
\right\},
\end{eqnarray}
\begin{eqnarray}
\label{55} {\cal T}(t)= \frac{\Theta (t)[a_{\cal T}^2+ b_{\cal
T}^2+2a_{\cal T}b_{\cal T}\cos(\Delta\omega t+\kappa)]}
{(\Delta\omega)^2+(\Gamma-\gamma_l)^2/4}\nonumber\\
+\frac{(1-\Theta(t))e^{\gamma_lt}[(\Delta\omega)^2+(\gamma+\gamma_l)^2/4]}
{(\Delta\omega)^2+(\Gamma+\gamma_l)^2/4},
\end{eqnarray}
where the designations are introduced:
\begin{equation}
\label{56} a_{\cal R}=e^{-\gamma_lt/2},\qquad b_{\cal
R}=\frac{\gamma_le^{-\Gamma t/2}}
{(\Delta\omega)^2+(\Gamma+\gamma_l)^2/4},
\end{equation}
\begin{eqnarray}
\label{57} a_{\cal
T}=\sqrt{(\Delta\omega)^2+(\gamma_l-\gamma)^2/4},\nonumber\\
b_{\cal
T}=\gamma_r\gamma_l/[2\sqrt{(\Delta\omega)^2+(\Gamma+\gamma_l)^2/4}],
\end{eqnarray}
 The angles  $\chi$, $\zeta$ and $\kappa$ are defined by the
expressions:
\begin{eqnarray}
\label{58} \cos\chi=-\frac{(\Delta\omega)^2+(\gamma_l+\Gamma)
+(\gamma_l+\gamma_r-\gamma)/4}
{\sqrt{[(\Delta\omega)^2+(\gamma_l+\Gamma)^2/4]}}\nonumber\\
\times\frac{1}
{\sqrt{[(\Delta\omega)^2+(\gamma_l+\gamma_r-\gamma)^2/4]}},\nonumber\\
\sin\chi=\frac{\Delta\omega\gamma}{\sqrt{[(\Delta\omega)^2+
(\gamma_l+\Gamma)^2/4]}}\nonumber\\
\times\frac{1}{\sqrt{[(\Delta\omega)^2+(\gamma_l+\gamma_r-\gamma)^2/4]}},
\end{eqnarray}
\begin{eqnarray}
\label{59}
\cos\zeta=-\frac{\gamma_l+\Gamma}{2\sqrt{(\Delta\omega)^2+(\gamma_l
+\Gamma)^2/4]}},\nonumber\\
\sin\zeta=\frac{\Delta\omega}{\sqrt{(\Delta\omega)^2+(\gamma_l+\Gamma)^2/4]}},
\end{eqnarray}
\begin{eqnarray}
\label{60}
\cos\kappa=-\frac{(\Delta\omega)^2+(\gamma_l-\gamma)(\gamma_l+\Gamma)/4}
{\sqrt{[(\Delta\omega)^2+(\gamma_l +\Gamma)^2/4]}}\nonumber\\
\times\frac{1}{\sqrt{[(\Delta\omega)^2+(\gamma_l-\gamma)^2/4]}},\nonumber\\
\sin\kappa=\frac{\Delta\omega(\Gamma+\gamma)/2}
{\sqrt{[(\Delta\omega)^2+(\gamma_l +\Gamma)^2/4]}}\nonumber\\
\times\frac{1}{\sqrt{[(\Delta\omega)^2+(\gamma_l-\gamma)^2/4]}}.
\end{eqnarray}
All the three dependencies contain the oscillations on the
frequency $\Delta\omega$ with different phase shifts.

\section{Specific  points for the detuned carrier frequency.}

Let us show that in the case $\Delta\omega\neq0$ the specific time
points of the first and second types do not exist,generally
speaking. Indeed, from the expressions Eq. (9),
 (10) and (15) it follows that the fields  $\Delta{\bf E}(z=0,t)$
 and ${\bf E}_{right}(z=0,t)$ may be written as follows
\begin{eqnarray}
\label{61} {\bf E}(z=0,t)=E_0{\bf
e}_l\{\Theta(t)[Ae^{-\gamma_lt/2-i(\omega_lt+\xi)}\nonumber\\+
Be^{-\Gamma t/2-i(\omega_0t+\varphi)}]+\nonumber\\ +[1-\Theta
(t)]Ce^{\gamma_lt/2-i(\omega_lt+\zeta)}\}+ c. c.,
\end{eqnarray}
where $A, B, C$ are the time independent real coefficients;
 $\xi$, $\varphi$ and $\zeta$ are the phase shifts. Using the
 circulate polarization Eq. (11), we transform Eq. (61) to the
expression
\begin{eqnarray}
\label{62} {\bf E}(z=0,t)=E_0\{\Theta (t)\nonumber\\
\times[A{\sqrt 2}e^{-\gamma_lt/2}({\bf e}_x
\cos(\omega_lt+\xi)\pm{\bf e}_y\sin(\omega_lt+\xi))]+\nonumber\\
+B\sqrt{2}e^{-\Gamma t/2}({\bf e}_x\cos(\omega_0t+\varphi)\pm{\bf
e}_y \sin(\omega_0t+\varphi))+\nonumber\\ +[1-\Theta
(t)]C\sqrt{2}e^{\gamma_lt/2}\nonumber\\ \times({\bf
e}_x\cos(\omega_0t+\zeta) \pm{\bf e}_y\sin(\omega_0t+\zeta))\},
\end{eqnarray}
where the sign  $+ (-)$ relates to the right (left) circular
polarization. Thus, at the detuning at $t>0$
 two circular polarized waves with the different carrier
frequencies  $\omega_l$ and $\omega_o$ and the different phase
shifts relative to ${\bf E}_0(z=0,t)$ are present in the
transmitting and reflected pulses. In such a case the summary
vector ${\bf E}_0(z=0,t)$ does not vanish, generally speaking.
Equating ${\bf E}_0(z=0,t)$ to 0, we obtain two equations  for one
unknown
 $t$, which have no solution. The same relates to the vector
 combinations
 ${\bf E}_0-\Delta{\bf E}$ and ${\bf E}_0+{\bf E}_{right}$.
Thus, the first and second types specific point  are absent in the
case of detuning. But the existence of the third type specific
points is possible, because the vectors $\Delta{\bf E}$, ${\bf
E}_{right}$, ${\bf E}_0-\Delta{\bf E}$ and ${\bf E}_0+{\bf
E}_{right}$ may be perpendicular to each other at some time
moments.

The conditions of existence of the zero absorption point, total
reflection and total transparency may be analysed conveniently
with the help of Eqs. (53) - (55). Below we confine ourselves with
analysis of the most interesting case $\gamma_r>>\gamma$, applying
$\gamma=0$. Then the condition $$\int_{-\infty}^{\infty}{\cal
A}_{\gamma =0}(t)dt=0$$ is fulfilled and it is clear, that at
least one zero absorption point exists. The analysis of Eq. (53)
at $\gamma =0$ shows that at $\gamma_l\neq\gamma_r$ the finite odd
number of the zero absorption points exist. That shows, that at
the large times
 ${\cal A}_{\gamma=0}(t)$
 becomes negative. The quantity of points,
where ${\cal A}_{\gamma=0}(t)=0,$ depends on the relationship
$q=\Delta\omega/\gamma_l$, i. e. on the detuning. In the case of
the short pulse $(\gamma_l>>\gamma_r)$ at $q<<\pi$ there exists
one point of zero absorption; many such points exist at $q>>\pi$.
In the case of a long pulse $(\gamma_l<<\gamma_r)$ at
 $q<<\pi\gamma_r/\gamma_l$ there is one zero absorption point;
 many such points exist at $q>>\pi\gamma_r/\gamma_l.$

In the case  $\gamma_l=\gamma_r$ there is the infinite number of
the zero absorption points. The quantity of the total reflection
points depends on the parameter $q$ . At $q<q_b$ there is the
infinite number of the total reflection points; at $q>q_b$ their
number equals to 0. The value $q_b$ is determined by the equation
$$(2q_b-1)\sqrt{1+q_b^2}-1=0$$ and equals $q_b=0.876$. Note also,
that at $\gamma_l=\gamma_r$ the number of the total transparency
points is infinite at any  $q$.

\section{Discussion of the detuning results.}

  The functions ${\cal R}(t)$, ${\cal A}(t)$ and ${\cal
T}(t)$ at $\gamma =0$  and the different interrelations of the
parameters $\gamma_r,$ $\gamma_l$
 and $\Delta\omega$ are represented in Figs. 4 - 6.
 Fig. 4 is relevant to the short pulse case
($\gamma_l>>\gamma_r$). Fig. 5 is relevant to the long pulse case
 ($(\gamma_l<<\gamma_r)$). Fig. 6 is relevant to the intermediate pulse
 case ($\gamma_l/\gamma_r=1$).

The dependencies of the reflection ${\cal R}(t)$ and the
absorption ${\cal A}(t)$ on the non-dimensional variable
$\gamma_lt$ at $\gamma_l/\gamma_r=10$ and for different values of
the parameter  
 $q=\Delta\omega/\gamma_l,$ characterizing detuning
, are represented in Fig. 4.
 For the short pulses the transmission curves ${\cal T}(t)$
are absent, because they are close to $P(t)$ for the values of $q$
from 1 to 10, i. e. transmission is always large. It is seen in
Fig. 4 that ${\cal R}$ and ${\cal A}$ are much smaller than 1 at
$\Delta\omega=0$ and decrease quickly with the increasing of the
detuning. The oscillating contribution into ${\cal R}$ and ${\cal
A}$ damps as $\exp[-(\gamma_l+\gamma_r)t/2]$. The damping
parameter
 is determined as $$\Pi=2/(1+\gamma_r/\gamma_l);$$ in Fig.
4 $\Pi=1,9$. The oscillation period $T=2\pi/q$. The oscillations
are absent on the curves ${\cal R}(t)$ and
 ${\cal A}(t)$ at $q=0,2$ because $T>>\Pi$.
 These oscillations are clearly seen on the curves  
where $q=1$ and $q=10$. With growth of the detuning the
oscillation period becomes smaller and the oscillation amplitude
decreases. The third type total reflection points are seen in Fig.
4a; the third type zero absorption  points are seen in Fig. 4b
 (see section IX).

The curves ${\cal R}$, ${\cal A}$ and ${\cal T}$ at
 $\gamma_l/\gamma_r=0,1$ are represented in Fig. 5.
 The oscillations are absent on the curves  ${\cal R}$ of Fig. 5a
  for any values $q$ because their amplitude is small in comparison
to the non-oscillating contribution. At $q=0$ the curve ${\cal
R}(t)$ is close to the curve
 $P(t)$, i. e. the pulse is almost totally reflected, but with growing $q$
 reflection decreases, reaching the very small values at $q=100$.
It is seen in Fig.5b , that absorption decreases sharply also with
growing parameter $q$.

The oscillations on the curves  ${\cal A}(t)$ are seen perfectly
at $q=20$ and $q=100$. The zero absorption points appear only at
$q=100$, which is in agreement with the results of section IX.
Transmission ${\cal T}$ (Fig. 5c) is small at $q=0$, it increases
with growing  $q$ and approaches  $P(t)$ at $q>10$. The
oscillations are absent. At $q=0$ there is the first type total
reflection point
 (Fig. 5 ), to which the zero absorption point in Fig. 5b and zero
 transparency point in Fig. 5c corresponds.
With growing $q$ the zero transparency point turns into the
minimum transmission point, and the total reflection point turns
into the maximal reflection point.

 Finally, the dependencies  ${\cal R}(t)$, ${\cal A}(t)$
and ${\cal T}(t)$ are represented in Fig. 6 for the intermediate
pulse ($\gamma_l=\gamma_r$). The curves ${\cal R}(t)$ at $q$ = 0,
1, 10 and 30 are represented in Fig. 6a. At $q=0$
   reflection is comparable with $P(t)$, but at $q=10$
 it becomes very small in comparison to unity.
 The oscillations are seen at $q=10$ and $q=30$.
 The first type total reflection point is seen
 on the curve $q=0,$
on the rest curves the total reflection points are absent. This is
in agreement with the result of section IX at $q>q_b$. The curves
${\cal R}\exp(\gamma_lt)$ are represented in Fig. 6b; the infinite
number of oscillations is seen due to excluded damping. In
accordance with section IX, there is an infinite number of the
third type total reflection points $q>q_b$; there are no total
reflection points at  $q<q_b.$ At $q=q_b$ the curve ${\cal
R}\exp(\gamma_lt)$ touches the curve
 $P(t)\exp(\gamma_lt)$, corresponding to the exciting pulse.
The curves ${\cal A}(t)$ for the intermediate pulse are
represented in Fig. 6c. With growing parameter $q$ absorption
decreases; the oscillations
 and a lot of the zero absorption points are clearly seen
 at $q=10$. Let us remember, that in the case of  $\gamma_l=\gamma_r$
  the number of the zero
 absorption points is infinite.
The curves  ${\cal T}(t)$ for the case of the intermediate pulse
are represented in Fig. 6d. With growing $q$ transmission
increases and at $q=2$ already approaches the curve $P(t)$. The
number of the total transparency points must be infinite at
 $q\ge 0$. However, due to the very fast damping of the curves
 only one total transparency point  is seen at $q=0,5$
 and two such points are seen at  
$q=2.$ In order to show a large number of the total transparency,
total reflection and zero absorption points, the curves
 ${\cal R}\exp(\gamma_lt)$,
 ${\cal A}\exp(\gamma_lt)$ and ${\cal T}\exp(\gamma_lt)$
 , corresponding to $q=0,7$, and the direct line $P\exp(\gamma_lt)$
 at $t>0$ are represented in Fig. 6e.
 It is seen that all three curves have different phase shifts.
 Let us  stress that in Figs. 6b and 6e
the curve ${\cal R}(t)$ never turns 0, as it may seem, but has
very small positive value in the minima.

\section{Conclusion.}

Thus, one can make the main qualitative conclusions, obtained
under condition $\gamma<<\gamma_r$. (For the sake of simplicity we
suppose $\gamma=0.$) In the resonance $\omega_l=\omega_0$ in the
case of a short pulse ($\gamma_l>>\gamma_r$), the pulse transmits
the QW almost without changing its profile. Reflection and
absorption are small, reflection is much smaller than absorption.
The very long pulse ($\gamma_l<<\gamma_r$) is reflected almost
completely, transmission is much smaller than absorption. Because
at $\gamma=0$ the integral absorption equals 0 at any
interrelations of
 $ \gamma_l$ and $\gamma_r$, there is the zero absorption point, where
 light energy absorption is alternated by its radiation.
 The total reflection happens  in the
 zero absorption point and transmission equals 0
 (the first type specific point).

The case $\omega_l=\omega_0, \gamma_l\simeq\gamma_r$ is of special
interest, where reflection, absorption and transmission are
comparable in values. The transmitting pulse profile distinguishes
drastically on the exciting pulse profile. Due to the presence of
the first type specific point, the transmitting pulse has two
maxima, i. e. it is two-hump-shaped.

 Under detuning
$\Delta\omega=\omega_l-\omega_0\neq0$, with growing deviation
$\Delta\omega$ reflection and absorption decrease and the
transmitting pulse approaches in value and profile to the exciting
pulse. At $\Delta\omega>>\gamma_l$ the pulse transmits the QW
almost unchanged. There are  oscillations  on the frequency
 $\Delta\omega$
 on all the curves ${\cal R}(t), {\cal A}(t)$ and ${\cal
T}(t).$ However, these oscillations are not seen in all Figs. 4 -
6 relevant to the detuning. At $\Delta\omega<<\gamma_l$ the
oscillation period is much longer than the time, at which the
damping of the values
 ${\cal R}(t), {\cal A}(t)$ and ${\cal T}(t)$ happens.
 At $\Delta\omega>>\gamma_l$
the oscillation period is small, their amplitude  is small also.
Therefore the oscillations are seen best of all at
$\Delta\omega\simeq\gamma_l$.

At the detuning the total reflection points, the total
transmission points and the zero absorption points do not coincide
with each other. These specific points were defined as the third
type points. In order to examine these points in the most
interesting case $\gamma_r\simeq\gamma_l$, Figs. 6b, 6e are
represented, where damping is omitted, because the values
 ${\cal R}(t), {\cal A}(t)$ and ${\cal T}(t)$,
 multiplied by $exp(\gamma_lt)$, are put on the ordinate axis.
It has been shown that in the case $\gamma_r=\gamma_l$ the number
of the zero absorption points and total transparency points is
infinite at
 $\Delta\omega\neq 0$,  and
 the number
of the total reflection points is infinite at
$\Delta\omega/\gamma_l>q_b$, where $q_b= 0,876$. The infinite
number of the zero absorption points means that the infinite
number of the energy transitions from the electronic system to the
light wave and vice verse happens, however, one has to remember
that these oscillations damp as $\exp(-\gamma_lt)$.

At the deviation from the equality $\gamma_r=\gamma_l$ the number
of the third type specific points is always finite. In particular,
the number of the zero absorption points is odd, because
absorption on the large times is always negative: the system gives
back the accumulated energy .

\section{Acknowledgements}
 This work has been partially supported by the Russian
Foundation for Basic Research (00-02-16904, 99-02-16628) and by
the Program "Solid State Nanostructures Physics".
 S.T.P thanks the Zacatecas Autonomous University and the National
Council of Science and Technology (CONACyT) of Mexico for the
financial support and hospitality. D.A.C.S. thanks CONACyT
(27736-E) for the financial support.
       Authors are grateful to A. D'Amore for a critical reading of the
manuscript.

\begin{figure}
\caption{Non-dimensional reflection
 ${\cal R}(t)$, absorption ${\cal A}(t)$ and transmission
 ${\cal T}(t)$ of the symmetrical light pulse by the two-level
 system. The case of the resonance
 $\Delta\omega=\omega_l-\omega_0=0$ and of the short pulse.
  $t_0$ is the first type total reflection point, $t_t$
  is the second type total transparency point. $\gamma_r$
  is the radiative inverse lifetime of the excited energy level,
  $\gamma$ is the non-radiative lifetime broadening
  of the excited energy level, $\gamma_l$
 determines the duration of the chosen
symmetrical pulse, $\omega_l$ is the pulse carrier frequency,
$\omega_0$ is the resonant transition frequency.}
\end{figure}
\begin{figure}
 \caption{ Same
as Fig. 1 for the long pulse. The total transparency point is
absent.}
\end{figure}
\begin{figure}
 \caption{ Same
as Fig~1 for the intermediate pulse.}
\end{figure}
\begin{figure}
\caption{The curves ${\cal R}(t)$ (a) and ${\cal A}(t)$ (b) in the
case of the very short pulse at different detuning values
 $\Delta\omega$.}
\end{figure}
\begin{figure}
\caption{ The curves ${\cal R}(t)$ (a), ${\cal A}(t)$ (b) and
${\cal T}(t)$ (c) in the case of the very long pulse at the
different detuning values $\Delta\omega$.}
\end{figure}
\begin{figure}
\caption{The curves ${\cal R}(t)$, ${\cal A}(t)$ and ${\cal T}(t)$
for the intermediate pulse at different values of $\Delta\omega$.
a - ${\cal R}(t)$, b - the curves ${\cal R}(t)\exp(\gamma_lt)$ for
$q<q_b, q=q_b$ and $q>q_b$, c - ${\cal A}(t)$, d- ${\cal T}(t)$, e
-the curves ${\cal R}(t)\exp(\gamma_lt)$, ${\cal
A}(t)\exp(\gamma_lt)$ and ${\cal T}(t)\exp(\gamma_lt)$ at
$q=0,7$.}
 \end{figure}


\begin{references}
\bibitem[\ddag]{byline2}
In the case of the pulse irradiation of a QW and under condition
 $\gamma_r<<\gamma$ the lowest approximation of the perturbation
 theory is applicable only at the times
 $t<<\gamma_r^{-1}$, because at $t\geq\gamma_r^{-1}$
the intensity of the transmitted and reflected light damps as
 $\exp(-\gamma_rt).$
\end{references}
\end{document}